\documentclass[11pt,a4]{article}
\usepackage[dvipdfmx]{graphicx}
\usepackage{latexsym}
\usepackage{amsmath} 
\usepackage{amsthm}
\usepackage{amsfonts}

\newtheorem{proposition}{Proposition}[section]
\newtheorem{lemma}{Lemma}[section]

\newcommand{\hide}[1]{}

\newcommand{\CC}{{\cal C}}
\newcommand{\TT}{{\cal T}}

\newcommand{\SSS}{{\cal S}}

\newcommand{\XX}{{\cal X}}
\newcommand{\bags}{{\mathop{\rm bags}}}
\newcommand{\caps}{{\mathop{\rm caps}}}
\newcommand{\local}{{\mathop{\rm local}}}
\newcommand{\width}{{\mathop{\rm width}}}
\newcommand{\tw}{{\mathop{\rm tw}}}
\newcommand{\ctw}{{\mathop{\rm ctw}}}
\newcommand{\xtw}{{\mathop{\rm xtw}}}
\newcommand{\gap}{{\mathop{\rm gap}}}
\title{A heuristic use of dynamic programming to upperbound treewidth}
\author{Hisao Tamaki}
\begin{document}
\maketitle

\begin{abstract}
For a graph $G$, let $\Pi(G)$ denote the set of all potential maximal cliques
of $G$. For each subset $\Pi$ of $\Pi(G)$, let $\tw(G, \Pi)$ denote the
smallest $k$ such that there is a tree-decomposition of $G$ of width 
$k$ whose bags all belong to $\Pi$.
Bouchitt\'{e} and Todinca observed in 2001 that
$\tw(G, \Pi(G))$ is exactly the treewidth of $G$ and
developed a dynamic programming algorithm 
to compute it. Indeed, their algorithm can readily be 
applied to an arbitrary non-empty subset $\Pi$ of $\Pi(G)$ and
computes $\tw(G, \Pi)$, or reports that it is undefined,
in time $|\Pi||V(G)|^{O(1)}$.
This efficient tool for computing $\tw(G, \Pi)$ allows
us to conceive of an iterative improvement procedure
for treewidth upper bounds which maintains, 
as the current solution, a set of potential maximal cliques rather than
a tree-decomposition.

We design and implement an algorithm along this approach.
Experiments show that our algorithm vastly outperforms previously
implemented heuristic algorithms for treewidth.
\end{abstract}

\newpage
\section{Introduction}
Treewidth is one of the most fundamental graph parameters which
plays an essential role in the graph minor theory \cite{RS86,RS04} and
is an indispensable tool in designing graph algorithms (see, for example, a survey \cite{BK08}).
We denote the treewidth of graph $G$ by $\tw(G)$.  See Section~\ref{sec:prelim} for 
the definition.
Since many NP-hard graph problems are tractable on graphs of small treewidth,    
computing the treewidth of a given graph is an extremely important problem.
Naturally, this problem has been actively studied from  
both theoretical \cite{ACP87,Bod96,BFKKT12,FV12} and 
practical \cite{GD04,BK10,SH09,BJ14,Tamaki18,Tamaki19} viewpoints.

Recently, there have been some progresses \cite{Tamaki18,Tamaki19} 
in practical algorithms for treewidth and
the scope of graph instances for which the treewidth can be exactly computed
in practice is significantly extended.

In \cite{Tamaki19}, it is observed that a promising approach to
exact treewidth computation is to have separate components for
upper and lower bounds rather than a single optimization algorithm.
For most instances for which the exact treewidth can be computed in
practical time at all, a good heuristic algorithm quickly produces
an upper bound that is equal to the exact width. In such cases,
the most time-consuming part of the exact computation would be for
showing the matching lower bound. Although vastly unexplored,
computation aimed at such tight lower bounds may also benefit
from heuristic approaches.

Having separate components for upper and lower bounds is even
more beneficial for practical treewidth computation, since,
for most instances of practical interest, computing the exact treewidth is
hopeless but a reasonably small gap between upper and lower bounds might
be achievable.  

In this paper, we are concerned with the
upper bound part of the computation.
One of the most successful approaches for this purpose is
iterative improvements. In this approach, the algorithm maintains a
{\em current solution} that is the best tree-decomposition found so far,
or is a variant stemmed from such a solution.
In each iteration, the algorithm tries to find a variation of this 
tree-decomposition that has a smaller width. Musliu's algorithm \cite{Musliu08},
which has the best experimental results in the literature, as well as 
the first place submission to PACE 2017 competition \cite{PACE17} (
the heuristic treewidth track) 
due to Otsuka {\it et al} are both based on this approach.
Those algorithms work well for small to medium-size instances
but do not seem to scale well for large instances.
Musliu's experiments are limited to graphs of fewer than 
1000 hundred vertices and mostly fewer than 500 vertices.
The PACE 2017 submission by Otsuka {\it et al} also performs
well for graphs of similar scale but, for larger instances, 
significantly poorer than the second place submission due to Ben Strasser,
which is based on a top down decomposition approach.
This is not surprising, as the 
improvements in the iterations of those algorithms are sought within a rather
small neighborhood of the current solution and 
it is imaginable that the search is  
trapped around a local optimum rather easily. 

In this paper, we introduce a new approach to computing an
upper bound on treewidth based on a heuristic use of the
dynamic programming algorithm due to Bouchitt\'{e} and Todinca 
(BT algorithm henceforth), which is originally designed
to compute the exact treewidth based on the notion of
potential maximal cliques. We recast their results in the following 
manner to suit our purposes.

A vertex set $X$ of $G$ is a {\em potential maximal clique} of $G$ 
if there is a minimal chordal graph $H$ with $V(G) = V(H)$ and $E(G) \subseteq E(H)$
such that $X$ is a maximal clique of $H$.
Let $\Pi(G)$ denote the set of all potential maximal cliques of $G$.
For each subset $\Pi$ of $\Pi(G)$, let $\tw(G, \Pi)$ denote
the smallest $k$ such that there is a tree-decomposition of $G$ of width $k$
all of whose bags belong to $\Pi$.
Bouchitt\'{e} and Todinca observed that $\tw(G, \Pi(G))$
is exactly the treewidth $\tw(G)$ of $G$ and developed a 
dynamic programming algorithm to compute it \cite{BT01}.
Indeed, their dynamic programming algorithm, which we refer to as the BT algorithm,
is readily applicable to an arbitrary subset $\Pi$ of $\Pi(G)$ and
computes $\tw(G, \Pi)$ in time $|\Pi||V(G)|^{O(1)}$.

Given this efficient algorithm for computing $\tw(G, \Pi)$,
we may conceive of an iterative improvement procedure
for treewidth upper bounds, which maintains as the current
solution a set of potential maximal cliques rather than a
tree-decomposition.

We show the effectiveness of this approach by experimentally evaluating  
an implementation of the approach. More specifically,
we run our implementation on the test instances 
from the heuristic treewidth track of PACE 2017 and
compare the results with those by the winning submissions.  
For each instance $G$, let $\tw_{{\rm PACE}}(G)$ denote the best upper bound 
over those computed by all the submissions and let $\tw_{{\rm OURS}}(G)$ denote
the upper bound computed by our implementation using computational
resources similar to those in PACE 2017.
For every instance $G$ with $|V(G)| \leq 10^4$
and $\tw_{{\rm PACE}}(G) \leq 200$
(there are 59 such out of 100 competition instances),
we have $\tw_{{\rm OURS}}(G) \leq \tw_{{\rm PACE}}(G) + 1$;
we have $\tw_{{\rm OURS}}(G) \leq \tw_{{\rm PACE}}(G)$ for
56 of those; we have $\tw_{{\rm OURS}}(G) < \tw_{{\rm PACE}}(G)$ for
29 of those. Thus, the performance of our implementation
is siginificantly better than the instance-wise best of
the PACE submissions, except for on graphs with
very large number of vertices or with very large width.

\section{Preliminaries}
\label{sec:prelim}
In this paper, all graphs are simple, that is, without self-loops or
parallel edges. Let $G$ be a graph. We denote by $V(G)$ the vertex set
of $G$ and by $E(G)$ the edge set of $G$.
For each $v \in V(G)$, $N_G(v)$ denotes the set of neighbors of $v$ in $G$:
$N_G(v) = \{u \in V(G) \mid \{u, v\} \in E(G)\}$.
For $U \subseteq V(G)$, the {\em open neighborhood of $U$ in $G$}, denoted by
$N_G(U)$,  is the set of vertices adjacent to some vertex in $U$ but not
belonging to $U$ itself: $N_G(U) = (\bigcup_{v \in U} N_G(v)) \setminus U$.
The {\em closed neighborhood of $U$ in $G$}, denoted by $N_G[U]$, is defined
by $N_G[U] = U \cup N_G(U)$.  We also write $N_G[v]$ for $N_G[\{v\}] = N_G(v)
\cup \{v\}$. We denote by $G[U]$ the subgraph of $G$ induced by $U$: 
$V(G[U]) = U$ and $E(G[U]) = \{\{u, v\} \in E(G) \mid u, v \in U\}$.
For $U \subseteq V(G)$, we also use the abbreviation $G \setminus U$ for
$G[V(G) \setminus U]$.
In the above notation, as well as in the notation further introduced below,
we will often drop the subscript $G$ when the graph is clear from the context. 

We say that vertex set $C \subseteq V(G)$ is {\em connected in}
$G$ if, for every $u, v \in C$, there is a path in $G[C]$ between $u$
and $v$. It is a {\em connected component} or simply a {\em component} 
of $G$ if it is connected and is inclusion-wise maximal 
subject to this condition. 

A {\em tree-decomposition} of $G$ is a pair $(T, \XX)$ where $T$ is a tree
and $\XX$ is a family $\{X_i\}_{i \in V(T)}$ of vertex sets of $G$ such that
the following three conditions are satisfied. We call members of
$V(T)$ {\em nodes} of $T$ and each $X_i$ the {\em bag} at node $i$.   
\begin{enumerate}
  \item $\bigcup_{i \in V(T)} X_i = V(G)$.
  \item For each edge $\{u, v\} \in E(G)$, there is some $i \in V(T)$
  such that $u, v \in X_i$.
  \item For each $v \in V(G)$, the set of nodes $I_v = \{i \in V(T) \mid v \in
  X_i\}$ of $V(T)$ induces a subtree of $T$.
\end{enumerate}
We may assume that the bags $X_i$ and $X_j$ are distinct from each other
for $i \neq j$ and, under this assumption, we will regard
a tree-decomposition as a tree $T$ in which each node is a bag.
Let $\bags(T)$ denote the set of bags of $T$.

To facilitate measuring the progress of our iterative improvement procedure,
our definition of the width of a tree-decomposision is a refinement of the
conventional one. 
The {\em width} of the tree-decomposition $T$, denoted by $\width(T)$
is a pair $(k, f)$ where $k = \max\{|X| - 1 \mid X \in \bags(T)\}$ and
$f = |\{X \in \bags(T) \mid |X| = k + 1\}|$. In words, $k$ is
the conventional width of $T$ which is the size of the largest bag of 
$T$ minus one and $f$ is the number of largest bags of $T$.
We use the natural lexicographic order on widths: we have 
$(k_1, f_1) < (k_2, f_2)$ if and only if $k_1 < k_2$ or $k_1 = k_2$ and 
$f_1 < f_2$. 
The {\em treewidth} of $G$, denoted by $\tw(G)$ is the minimum width
of all tree-decompositions of $G$.
For each pair $w = (k, f)$, we denote by $\TT_w(G)$
the set of all tree-decompositions of $T$ of width at most $w$. 
A tree-decomposition $T$ of $G$ is {\em optimal} if the width of $T$ equals $\tw(G)$.

It turns out convenient, in combining subtrees into a tree-decomposition,
to define the addition of widths:
for $w_1 = (k_1, f_1)$ and $w_2 = (k_2, f_2)$, 
$w_1 + w_2$ is $(k_1, f_1 + f_2)$ if $k_1 = k_2$,
$w_1$ if $k_1 > k_2$, and $w_2$ if $k_1 < k_2$.
%For notational convenience we define,
%for $w = (k, f)$, that $w - 1 = (k, f - 1)$ if $f > 0$ and
%$w - 1 = (k - 1, n)$ for $f = 0$, where $n$ is the number of vertices of $G$.
%We note that the tree-decompositions our algorithm deals with never has
%more than $n$ bags and therefore the convention in the latter case is justified.

\section{Minimal triangulations and potential maximal cliques}

The close relationship between the tree-decompositions of a graph and the triangulations of a graph
plays an important role in many treewidth algorithms. We collect basic facts about this
relationship in this section.
The proofs can either be found in the literature \cite{Heggernes06, BT01} or
easily be inferred from the basic facts therein.  
%For self-containedness, we prove those facts in the appendix.

Graph $G$ is {\em chordal} if every induced cycle of $G$ has length exactly
three. A vertex set $S$ of $G$ is a {\em clique} of $G$ if $G[S]$ is a complete graph;
it is a {\em maximal clique} of $G$ if it is a clique of $G$ and no proper
superset of it is a clique.
For a chordal graph $G$, let $\Pi(G)$ denote the set of maximal cliques
of $G$. 

For each vertex set $S$ of $G$, let $\CC_G(S)$ denote
the set of all components of $G \setminus S$.
We call $S$ a {\em separator} of $G$ if $\CC_G(S)$ has
more than one members.
We call $C \in \CC_G(S)$ 
a {\em full component} of $S$ if $N(C) = S$.
A vertex set $S$ of $G$ is a {\em minimal separator}
if $\CC_G(S)$ has at least two full components of $S$.
Note that $S$ is a minimal separator
of $G$ if and only if there are two vertices $a, b$ of $G$ 
that belong to distinct components of $G \setminus S$ but
belongs to the same component of $G \setminus S'$ for every proper
subset of $S'$ of $S$. Note also that, if $G$ is disconnected,
then the empty set is a minimal separator of $G$. 

\begin{proposition}
Let $G$ be a chordal graph. Then, there is a tree-decomposition $T$ of $G$
such that $\bags(T) = \Pi(T)$. In such $T$, the intersection of
every pair of adjacent bags is a minimal separator of $G$.
\end{proposition}
We call the tree-decomposition in this proposition a {\em clique tree} of $G$.
Clearly, a clique tree of a chordal graph $G$ is an optimal tree-decomposition of 
$G$, since if $T$ is an arbitrary tree-decoposition of $G$, then, for each clique $K$ of $G$,
$T$ must have a bag $X$ with $K \leq X$. 

For a general graph $G$, graph $H$ is a {\em triangulation of $G$} if it is chordal,
$V(H) = V(G)$, and $E(G) \subseteq E(H)$. It is a {\em minimal triangulation}
of $G$ if moreover $E(H)$ is inclusion-wise minimal subject to these conditions.

We say that a separator $S$ of $G$ {\em crosses} a vertex set $U$ of
$G$ if at least two members of $\CC_G(S)$ intersects $U$.
Observe that, if $S_1$ and $S_2$ are minimal separator of $G$,
then $S_1$ crosses $S_2$ if and only if $S_2$ crosses $S_1$.
A set $\Delta$ of minimal separators of $G$ is {\em non-crossing}
if no two members of $\Delta$ crosses each other.

\begin{proposition}
\label{prop:non-crossing} 
Let $\Delta$ be a non-crossing set of minimal separators of $G$ and
let $G'$ be obtained from $G$ by filling every member of $\Delta$ into
a clique.  Then, every minimal triangulation of $G'$ is a minimal
triangulation of $G$.
\end{proposition}

\begin{proposition}
\label{prop:clique-minsep}
Let $S$ be a minimal separator of $G$ and suppose $S$ is a clique of
$G$. Let $H$ be an arbitrary minimal triangulation of 
$G$. Then the edge set of $H$ is the union of
edge sets of $H[N[C]]$, $C \in \CC_G(S)$.
\end{proposition}

A vertex set $S$ of $G$ is a {\em potential maximal clique} of $G$ if it
is a maximal clique of some minimal triangulation of $G$.
We denote by $\Pi(G)$ the set of all potential maximal cliques of $G$.
This is consistent with our notation for chordal graphs, 
since if $G$ is chordal then $G$ is the unique minimal triangulation of $G$ 
and the set of potential maximal
cliques coincides with the set of maximal cliques.

Let $S \subseteq V(G)$. We define $\SSS_G(S) = \{N(C) | C \in \CC_G(S)\}$.
The {\em local graph} of $G$ on $S$, 
denoted by $\local(G, S)$, is the graph obtained from
$G[S]$ by filling every $U \in \SSS_G(S)$ into a clique.

\begin{proposition}
Let $G$ be a graph. A vertex set $X$ of $G$ is a potential maximal clique of $G$ if and only if
the following conditions hold:
\begin{enumerate}
  \item no member of $\CC_G(X)$ is a full component of $S$, and 
  \item $\local(G, X)$ is a clique of $G$.
\end{enumerate}
Moreover, if $X$ is a potential maximal clique of $G$, 
then, each $S \in \SSS_G(X)$ is a
minimal separator of $G$.
\end{proposition}

For each minimally separated connected set $C$ of $G$, 
we say that $X \in \Pi(G)$ is a
{\em cap} of $C$ if $N(C) \subseteq X \subseteq N[C]$.

\begin{proposition}
Let $X \in \Pi(G)$ and $S \in \SSS_G(X)$.
Then, there is a unique minimally separated connected set $C$ of
$G$ such that $N(C) = S$ and $X$ is a cap of $C$.
\end{proposition}

For each minimally separated connected set $C$ of $G$, we denote
by $\caps_G(C)$ the set of all caps of $C$.

\begin{proposition}
\label{prop:local}
Let $G$ be a graph and $S$ an arbitrary vertex set of $G$.
Then, every potential maximal clique of $\local(G, S)$ is
a potential maximal clique of $G$ unless it is $N_G(C)$ for
some connected component $C$ of $G \setminus S$. 
\end{proposition}

For each subset $\Pi$ of $\Pi(G)$, let $\TT(G, \Pi)$ denote
the set of all tree-decompositions $T$ such that $\bags(T) \subseteq \Pi$.

The {\em treewidth of $G$ in $\Pi$},
denoted by $\tw(G, \Pi)$, is the smallest $w$ such that 
$\TT(G, \Pi) \cap \TT_w(G)$ is non-empty; 
if $\TT(G, \Pi)$ is empty then, $\tw(G, \Pi)$ is undefined.

\begin{proposition}
Let $G$ be a graph. Then, $\tw(G) = \tw(G, \Pi(G))$.
\end{proposition}

For $\Pi \subseteq \Pi(G)$ such that $w = \tw(G, \Pi)$ is defined,
the {\em core} of $\Pi$ is a minimal subset $\Pi'$ of $\Pi$
such that $\TT(G, \Pi') \cap \TT_{w}(G) = \TT(G, \Pi) \cap \TT_{w}(G)$.

We note that, given $G$ and $\Pi \subseteq \Pi(G)$ such that
$\tw(G, \Pi)$ is defined, the BT algorithm computes 
$\tw(G, \Pi)$ and the core of $\Pi$ in time $|\Pi||V(G)|^{O(1)}$.

\section{Component algorithms}
\label{sec:components}
\subsection{BT dynamic programming}
In this section, we review the dynamic programming algorithm
due to Bouchitt\'{e} and Todinca 
to compute $\tw(G, \Pi)$, given a graph $G$ and
a set $\Pi$ of potential maximal cliques of $G$.
We assume that the input graph $G$ is connected in the rest of this paper.

The following lemma states our version of the recurrence
used in the BT dynamic programming algorithm.
We say that a connected set $C$ of $G$ is {\em minimally separated}
if either $N(C)$ is a minimal separator or $C = V(G)$.
For each minimally separated connected set $C$ of $G$ and a set of
potential maximal cliques $\Pi$ of $G$, the {\em component treewidth
of $C$ in $G$ with respect to $\Pi$}, denoted by
$\ctw(G, C, \Pi)$, is defined to be $\tw(\local(G, N[C]), \Pi')$,
where $\Pi'$ is the set of all potential potential maximal
cliques of $\local(G, N[C])$ that belong to $\Pi$.
If $\tw(\local(G, N[C]), \Pi')$ is undefined, then $\ctw(G, C, \Pi)$
is also undefined. 

\begin{lemma}
\label{lem:recurrence}
Let $G$ be a graph and $\Pi \subseteq \Pi(G)$.
Let $C$ be a minimally separated connected set
of $G$. Then, we have 
\begin{eqnarray*}
\ctw(G, C, \Pi) = \min_{X \in \caps_G(C) \cap \Pi}
((|X| - 1, 1) + \sum_{D \subseteq C, N(D) \subseteq X} \ctw(G, D, \Pi)), 
\end{eqnarray*}
where the addition of widths are as defined in the previous section.
It is understood that the minimum is taken over all $X$ for which
all of the terms in the summation are defined: if no such $X$ exists,
then $\tw(G, C, \Pi)$ is undefined.
\end{lemma}

The dynamic programming algorithm goes as follows.
We are given a graph $G$ and $\Pi \subseteq \Pi(G)$.
\begin{enumerate}
  \item Compute $\CC = \{V(G)\} \cup \bigcup_{X \in \Pi} \CC_G(X)$.
  \item Compute $\caps(C, \Pi) = \caps_G(C) \cap \Pi$ for each
  $C \in \CC$ as follows: for each $X \in \Pi$ and each $S \in \SSS_G(X)$,
  let $D$ be the unique full component of $S$ that contains $X \setminus S$;
  if $D \in \CC$ then put $X$ in $\caps(D, \Pi)$.
  \item Scan the members of $\CC$ in the ascending order of the cardinality and
  compute $\ctw(G, C, \Pi)$ for each $C \in \CC$ using the recurrence in Lemma~\ref{lem:recurrence}.
  \item Conclude that $\tw(G, \Pi) = \ctw(G, V(G), \Pi)$ if the right hand size
  is defined. Otherwise, $\tw(G, \Pi)$ is undefined.
\end{enumerate}

%\begin{theorem}
%The above algorithm correctly computes $\tw(G, \Pi)$.
%\end{theorem}
%\begin{proof}
%\end{proof}

\subsection{Minimal triangulation}
Another important sub-algorithm is the algorithm for minimal triangulation
of graphs. This sub-algorithm is applied to various local graphs of the given graph
and generates potential maximal cliques of the local graphs, 
which are candidates of potential 
maximal cliques of the entire graph due to Proposition~\ref{prop:local}.
 
Our algorithm for this purpose, which we call MMAF, is a variant of the algorithm 
due to Berry, Heggernes and Simonet \cite{BHS2003} called MMD.
MMD is based on the well-known minimum degree heuristic MD
for triangulating graphs. MMD uses MD as a subprocedure and,
unlike MD, returns a triangulation that is guaranteed to be minimal. Our variant, which we call
MMAF, replaces MD by another greedy heuristic MAF (minimum average-fill)
in MMD. It is observed \cite{OKST18} that
the triangulation MAF returns usually has smaller maximum clique size 
than the triangulation MD returns. We turn MAF into MMAF to return 
minimal triangulations, applying the method of the above authors which turns MD into MMD.

\section{Main algorithm}
In this section, we describe our main algorithm that computes an upper bound
on the treewidth of the given graph $G$. We first outline the algorithm.
\begin{enumerate}
  \item Use MMAF to compute a minimal triangulation $H$ of $G$.
  \item Let $\Pi_0 = \Pi(H)$ and
  let $w_0$ be the width of a clique-tree of $H$, which 
  is an invariant over all clique-trees of $H$, since they have the same set of bags, 
  namely $\Pi_0$.
  \item Repeat the following for $i = 1, 2, \ldots$, maintaining the invariant
  $\tw(G, \Pi_i) \leq w_i$ and $w_i < w_{i - 1}$ as long as the $i$th iteration
  is completed.
  \begin{enumerate}
    \item Let $\Pi = \Pi_{i - 1}$.
    \item Repeatedly add some potential maximal cliques to $\Pi$,
    using the methods to be described below, until
    $\tw(G, \Pi) < w_{i - 1}$.
    \item Set $w_i$ to $\tw(G, \Pi)$.
    \item Set $\Pi_i$ to the core of $\Pi$.
  \end{enumerate}
\end{enumerate}

In step 3(b) above, we look for potential maximal cliques to be added
to $\Pi$ so that $\tw(G, \Pi)$ decreases after accumulating them.
Once we have $\tw(G, \Pi)$ decreased, we shrink $\Pi$ by taking its core
in step 3(d).

Adding potential maximal cliques is done with two major strategies:
diversification and connection.  
Diversification is meant to add some potential maximal cliques to $\Pi$
that are ``essentially different'' from those in $\Pi$, in the
sense that they have the potential of supporting tree-decompositions
not similar to the ones in $\TT(G, \Pi)$. 
Connection is meant to add potential maximal cliques to $\Pi$ 
that are potentially used to connect up partial tree-decompositions
already possible with bags of $\Pi$ into a complete tree-decomposition
or a larger partial tree-decomposition. More details of these
two strategies are given below. 

\subsection{Diversification}
Let $\Pi \subseteq \Pi(G)$ is given such that
$\tw(G, \Pi) = (k, f)$ is defined.
Let $\hat{\Pi}$ be the core of $\Pi$.
For diversification, we pick a random element $X_0$ of $\hat{\Pi}$
with $|X_0| = k$
and a random subtree $R$ of a random tree-decomposition $T$ in
$\TT(G, \hat{\Pi})$, such that $X_0 \in \bags(R)$. 
Let $U = \bigcup_{X \in \bags(R)} X$ and $H = \local(G, U)$.
Observe that $R$ is a tree-decomposition of $H$.
Observe also that if we have some tree-decomposition
$R'$ of $H$ with $\width(R') < \width(R)$ then,
replacing $R$ by $R'$ in $T$, we obtain a tree-decomposition
$T'$ of $G$ with $\width(T') < \width(T)$.
With these observations in mind, we 
compute several minimal triangulations of $H$ 
as described below and, 
for each such minimal triangulation $H'$, 
add each members of $\Pi(H')$ that is a potential maximal clique of
$G$ to $\Pi$.
If $\tw(H') < \width(R)$ then, as observed as above, 
we have $\tw(G, \Pi \cup \Pi(H')) < \tw(G, \Pi)$ and
hence we have achieved the goal of improving $\Pi$.
Even if this is not the case, we may expected that
the added potential maximal cliques are useful in lowering
$\tw(G, \Pi)$, possibly together with future additions.

The method of computing several minimal triangulations of
$H$ is as follows. For diversity, we do not want $X_0$
to be a maximal clique in any of those minimal triangulations.
To ensure this, we first list several minimal separators
$S_1$, \ldots, $S_m$ of $H$ such that each $S_i$ crosses
$X_0$.
For $i = 1$, \ldots, $m$,
we let $H_i$ be obtained from $H$ by filling $S_i$ into a clique
and then apply MMAF to $H_i$ to obtain a minimal triangulation
$H_i'$ of $H_i$. By Proposition~\ref{prop:non-crossing},
$H_i'$ is a minimal triangulation of $H$, in which 
$X_0$ is not a maximal clique of $H_i'$ due to 
Proposition~\ref{prop:clique-minsep}.

\subsection{Connection}
Suppose the current set of potential maximal cliques is $\Pi$ and 
let $\tw(G, \Pi) = w = (k, f)$.
Denote by $\CC_G(\Pi)$ the union $\bigcup_{X \in \Pi} \CC_G(X)$.
For a pair $C, D \in \CC_G(\Pi)$, where $D$ is a proper subset of $C$, 
define the {\em external width} of $(C, D)$ with respect to $\Pi$, 
denoted by $\xtw(C, D, \Pi)$,  to be 
$\tw(G, D, \Pi) + \sum_{C' \in \CC_G(N(C)) \setminus \{C\}}  
\tw(G, C', \Pi)$.  The {\em gap} of this pair,
denoted by $\gap(C, D)$, 
is the local graph $\local(G, U)$ on $U$, where $U = N[C] \setminus D$.
We say that this pair 
is {\em promising} if $\xtw(C, D, \Pi) < w$.
Observe that if we find a minimal triangulation $H$ of $\gap(C, D)$
with maximum clique size at most $k$ 
for a promising pair $(C, D)$, then we have 
$\tw(G, \Pi \cup \Pi(H)) < w$ and we are successful in 
lowering the treewidth. We say that the set $\Pi(H)$
{\em fulfills} the promising pair if this is the case.
We say that a set of potential maximal cliques $\Delta \Pi$ 
{\em connects} this pair if there is a minimal triangulation $H$ of
$\gap(C, D)$ such that $\Delta \Pi$ forms a path
in a clique tree of $H$ between some $X_C$ with $N(C) \subseteq X$ and
some $X_D$ with $N(D) \subseteq X_D$, and 
the largest member of $\Delta \Pi$ has cardinality at most $k$.

The connection strategy in general picks up a
promising pair and tries to find the set of potential 
maximal cliques that fulfills or connects the pair. 
When successful in finding a fulfilling set,
enriched $\Pi$ with this set immediately lowers
$\tw(G, \Pi)$. Finding a connecting set is also considered
a major progress towards lowering the treewidth.
When we are not successful in finding a set that is fulfilling or
connecting, we still add some potential maximal cliques found in the
process, in hope that accumulating them will eventually
lead to an improved width.

\subsubsection{Direct connection}
Given a promising pair $(C, D)$, we look for
a singleton set $\{X\}$ that connects this pair.
In other words, $X$ is a cap of $C$ with $|X| \leq k$ such that $D \in \CC_G(X)$.
To this end, let $S = N(C) \cup N(D)$. First suppose that
$\CC_G(S)$ does not contain a full component of $S$.
If, moreover, $\local(G, S)$ is a clique, then 
$S$ is a potential maximal clique. So we add $S$ to $\Pi$
if $|S| \leq  k$. If $\local(G, S)$ is not a clique, 
we apply our greedy heuristic to obtain a minimal triangulation
$H$ of $\local(G, S)$. We add each member of $\Pi(H)$ with
cardinality at most $k$ that is a potential maximal clique of 
$G$ to $\Pi$. In either case, we probably have not found a set of potential maximal cliques
that fulfills the promising pair $(C, D)$, but possibly have
found some subset of a fulfilling set.

\subsubsection{Greedy remote connection}
Given a promising pair $(C, D)$, 
we use our greedy heuristic
to compute a minimal triangulation $H$ of
$\gap(C, D)$. If the maximum clique size of $H$ is
$k$ or smaller, then $\Pi(H)$ fulfills the promising pair.
Even if we do not have this luck, we 
add each member of $\Pi(H)$ with cardinality 
at most $k + 1$ that is a potential maximal clique of $G$ to $\Pi$.  

\subsubsection{Path connection}
Given a promising pair $(C, D)$, 
we look for a sequence $X_0$, $X_1$, \ldots, $X_d$ of
potential maximal cliques and a sequence 
$C_0$, $C_1$, \ldots, $C_d$ of connected sets such that
\begin{enumerate}
  \item $D \subseteq C_d \subseteq \ldots C_1 \subseteq C_0 = C$,
  \item $X_i$ is a cap of $C_i$ for $1 \leq i \leq d$,
  \item $C_i$ is the only member of $\CC_G(X_{i - 1})$ 
  that is also a subset of $C_{i - 1}$, for $0 \leq i \leq d$, and
  \item $|X_i| < k + 1$ for $1 \leq i \leq d$.
\end{enumerate}
If such a sequence is found, we add $X_0$, $X_1$, \ldots, $X_d$ to
$\Pi$. If $C_d$ happens to be $D$, then this added set 
connects the promising pair $(C, D)$ and indeed, due to the conditnion 3 above,
fulfills this pair.

We use depth-first search to find such a sequence and, 
to control the time for the search, we limit the length $d$ of
the sequence by a certain constant. In the current implementation,
this limit is set to 20.

\section{Experimental results}
\label{sec:experiments}
This section reports the result of the experiment we performed.
Test instances are taken from PACE 2017 \cite{PACE17} heuristic treewidth track
and the upper bounds computed by our algorithm are compared to
those computed by the top three submissions of PACE. 
Out of the 100 instances used for ranking the submissions, 
we take 83 instances with the number of vertices at most $100000$.

As in the PACE 2017 competition, the timeout of 30 minutes for each instance
is used for our computation.

The computing environment for our experiments is as follows.
CPU: Intel Core i7-8700, 3.20GHz; RAM: 32GB; 
Operating system: Windows 10, 64bit; 
Programming language: Java 1.8; JVM: jre1.8.0\_201.
The maximum heap size is set to 28GB. The implementation is single threaded, 
except that multiple threads may be invoked for garbage collection by JVM.

In the PACE 2017 competition, the following server was used 
for evaluating submissions. 
Dell PowerEdge R920;  
CPU: 4 x 3.0GHz Intel Xeon E7-8857 v2; Main memory: 1.5 TB 
Operating system: Debian jessie with linux 4.4.30.1.amd64-smp.

It is hard to compare the relative computing speeds of these two platforms but,
for single thread tasks, they are likely to be reasonably comparable.
Also note that our comparisons based on the upper bounds computed are
relatively insensitive to the computing time. In most cases, for example, 
doubling the timeout would change the results only slightly, if at all. 

Tables~\ref{tab:PACE1} and \ref{tab:PACE2} show the
result of comparisons. Each row consists of
the instance name, the number of vertices, the number of edges,
followed by the upper bounds on treewidth:
best of all PACE submissions, by the first place submission,
by the second place submission, by the third place submission, 
and by our algorithm HBT (for heuristic BT).
The instances are sorted in the increasing order of
the best of PACE treewidth upper bounds.
The results of the PACE winners are taken from the
file ``ranks-he.txt'' sent to the participants by the
PACE organizer.

\begin{table}[htbp]
  \begin{center}
    \begin{tabular}{|c|r|r|r|r|r|r|r|} \hline
    name & $|V|$& $|E|$ & bestPace & 1stPace & 2ndPace & 3rdPace& HBT\\
    \hline 
he010 & 82 & 146 & {\bf 5} & {\bf 5} & {\bf 5} & {\bf 5} & {\bf 5}\\
he006 & 111 & 199 & {\bf 7} & {\bf 7} & {\bf 7} & {\bf 7} & {\bf 7}\\
he002 & 172 & 408 & {\bf 9} & {\bf 9} & 10 & 10 & {\bf 9}\\
he004 & 172 & 408 & {\bf 9} & {\bf 9} & 10 & 10 & {\bf 9}\\
he054 & 51 & 240 & {\bf 11} & {\bf 11} & {\bf 11} & {\bf 11} & {\bf 11}\\
he008 & 332 & 580 & {\bf 13} & {\bf 13} & 14 & 14 & {\bf 13}\\
he076 & 90 & 135 & {\bf 15} & {\bf 15} & 20 & 17 & {\bf 15}\\
he038 & 604 & 1128 & 19 & 19 & 23 & 22 & {\bf 18}\\
he034 & 836 & 1436 & 20 & 20 & 22 & 22 & {\bf 19}\\
he058 & 112 & 168 & {\bf 21} & {\bf 21} & 23 & 22 & {\bf 21}\\
he056 & 112 & 168 & 23 & 24 & 24 & 23 & {\bf 22}\\
he020 & 849 & 1503 & {\bf 23} & {\bf 23} & 28 & 28 & 24\\
he062 & 126 & 189 & {\bf 24} & {\bf 24} & 28 & 26 & {\bf 24}\\
he052 & 193 & 642 & 24 & 24 & 26 & 27 & {\bf 23}\\
he012 & 921 & 1614 & 24 & 24 & 29 & 28 & {\bf 22}\\
he080 & 50 & 175 & {\bf 25} & {\bf 25} & 28 & 27 & {\bf 25}\\
he028 & 786 & 1455 & 25 & 25 & 27 & 26 & {\bf 22}\\
he122 & 7343 & 14352 & {\bf 25} & {\bf 25} & 30 & 30 & {\bf 25}\\
he032 & 676 & 1348 & 26 & 26 & 32 & 31 & {\bf 25}\\
he018 & 762 & 1472 & 28 & 28 & 33 & 32 & {\bf 25}\\
he024 & 773 & 1419 & 28 & 28 & 32 & 30 & {\bf 26}\\
he014 & 739 & 1451 & 29 & 29 & 34 & 33 & {\bf 26}\\
he036 & 817 & 1522 & 30 & 30 & 33 & 30 & {\bf 27}\\
he132 & 7565 & 14847 & {\bf 30} & {\bf 30} & 36 & 36 & 31\\
he030 & 661 & 1283 & 31 & 31 & 36 & 35 & {\bf 29}\\
he060 & 56 & 280 & {\bf 32} & {\bf 32} & 36 & {\bf 32} & {\bf 32}\\
he124 & 7420 & 14558 & {\bf 32} & {\bf 32} & 36 & 36 & 33\\
he068 & 112 & 336 & {\bf 33} & {\bf 33} & 40 & 42 & {\bf 33}\\
he104 & 1236 & 11416 & {\bf 33} & {\bf 33} & {\bf 33} & {\bf 33} & {\bf 33}\\
he098 & 708 & 3714 & {\bf 35} & {\bf 35} & 53 & 48 & {\bf 35}\\
he026 & 725 & 1397 & 35 & 35 & 38 & 36 & {\bf 30}\\
he022 & 735 & 1479 & 35 & 35 & 40 & 40 & {\bf 33}\\
he016 & 790 & 1523 & 39 & 43 & 40 & 39 & {\bf 34}\\
he042 & 179 & 2428 & {\bf 40} & {\bf 40} & 42 & 42 & {\bf 40}\\
he064 & 68 & 408 & {\bf 42} & {\bf 42} & {\bf 42} & {\bf 42} & {\bf 42}\\
he082 & 56 & 756 & {\bf 43} & {\bf 43} & {\bf 43} & {\bf 43} & {\bf 43}\\
he066 & 121 & 1265 & {\bf 50} & {\bf 50} & 54 & 54 & {\bf 50}\\
he050 & 192 & 1230 & {\bf 53} & 58 & {\bf 53} & {\bf 53} & {\bf 53}\\
he090 & 1558 & 2898 & 56 & 56 & 61 & 61 & {\bf 49}\\
he094 & 1684 & 3210 & 58 & 65 & 58 & 59 & {\bf 50}\\
he044 & 125 & 736 & 61 & 61 & 66 & 65 & {\bf 60}\\
    \hline 
\end{tabular}
    \caption{Results on PACE 2017 comppetition instances (1)}
    \label{tab:PACE1}
\end{center}
\end{table}

\begin{table}[htbp]
  \begin{center}
    \begin{tabular}{|c|r|r|r|r|r|r|r|} \hline
        name & $|V|$& $|E|$ & bestPace & 1stPace & 2ndPace & 3rdPace& HBT\\
    \hline 
he072 & 144 & 1750 & {\bf 61} & {\bf 61} & 66 & 71 & {\bf 61}\\
he070 & 85 & 850 & {\bf 63} & {\bf 63} & 64 & 64 & {\bf 63}\\
he040 & 196 & 4596 & {\bf 70} & {\bf 70} & 79 & 79 & {\bf 70}\\
he106 & 4660 & 12568 & {\bf 71} & 89 & 75 & {\bf 71} & {\bf 71}\\
he046 & 340 & 1425 & 80 & 80 & 91 & 91 & {\bf 79}\\
he096 & 1675 & 3425 & 80 & 88 & 80 & 89 & {\bf 64}\\
he084 & 2113 & 4373 & 92 & 114 & 92 & 94 & {\bf 86}\\
he092 & 1848 & 3574 & 94 & 102 & 94 & 94 & {\bf 78}\\
he108 & 1160 & 21206 & {\bf 97} & {\bf 97} & 111 & 114 & {\bf 97}\\
he086 & 1996 & 4007 & 97 & 116 & 97 & 99 & {\bf 79}\\
he088 & 2030 & 4215 & 98 & 110 & 98 & 114 & {\bf 82}\\
he048 & 363 & 2992 & 99 & 99 & 104 & 109 & {\bf 95}\\
he112 & 3228 & 25008 & 106 & 111 & 118 & 106 & {\bf 78}\\
he074 & 121 & 3630 & {\bf 108} & {\bf 108} & 109 & 109 & {\bf 108}\\
he078 & 125 & 6961 & {\bf 119} & {\bf 119} & {\bf 119} & {\bf 119} & {\bf 119}\\
he102 & 823 & 9422 & 122 & 122 & 150 & 149 & {\bf 114}\\
he150 & 31473 & 92564 & 124 & 132 & 192 & 124 & {\bf 99}\\
he110 & 1133 & 14629 & 184 & 184 & 212 & 193 & {\bf 183}\\
he140 & 20103 & 59932 & {\bf 216} & 20102 & {\bf 216} & 280 & 230\\
he138 & 17307 & 51995 & {\bf 218} & 17306 & {\bf 218} & 264 & 234\\
he130 & 2257 & 29919 & {\bf 230} & 461 & {\bf 230} & 257 & 248\\
he114 & 1178 & 25071 & {\bf 235} & 340 & {\bf 235} & 272 & 286\\
he120 & 1266 & 29837 & 235 & 394 & 235 & 272 & {\bf 205}\\
he134 & 1995 & 32272 & {\bf 239} & 483 & {\bf 239} & 304 & 337\\
he144 & 23695 & 66385 & {\bf 246} & 263 & {\bf 246} & 280 & 274\\
he100 & 500 & 62500 & {\bf 250} & {\bf 250} & {\bf 250} & {\bf 250} & {\bf 250}\\
he154 & 66421 & 192918 & {\bf 335} & 642 & {\bf 335} & 469 & 504\\
he152 & 72835 & 217600 & {\bf 354} & 418 & {\bf 354} & 411 & 413\\
he156 & 85646 & 227964 & 397 & 501 & 397 & 431 & {\bf 386}\\
he146 & 26328 & 71409 & {\bf 426} & 701 & {\bf 426} & 600 & 602\\
he158 & 6220 & 863026 & 483 & 623 & 527 & 483 & {\bf 391}\\
he148 & 34033 & 94695 & {\bf 501} & 692 & {\bf 501} & 572 & 612\\
he142 & 9600 & 94730 & {\bf 504} & 874 & {\bf 504} & 621 & 586\\
he164 & 20446 & 778291 & {\bf 557} & {\bf 557} & 558 & 558 & 558\\
he136 & 8591 & 34905 & {\bf 575} & 583 & 607 & {\bf 575} & {\bf 575}\\
he162 & 95695 & 336775 & {\bf 589} & 707 & {\bf 589} & 649 & 615\\
he160 & 62914 & 462976 & 672 & 672 & 744 & 672 & {\bf 654}\\
he126 & 2200 & 23538 & {\bf 869} & {\bf 869} & 932 & 951 & 873\\
he128 & 2200 & 23475 & 913 & 932 & 913 & 957 & {\bf 871}\\
he118 & 2200 & 23502 & 915 & 957 & 915 & 921 & {\bf 862}\\
he116 & 2200 & 23479 & 953 & 1002 & 955 & 953 & {\bf 892}\\
he190 & 2200 & 23479 & 4108 & 5068 & 4108 & 4647 & {\bf 3252}\\
    \hline 
\end{tabular}
    \caption{Results on PACE 2017 comppetition instances (2)}
    \label{tab:PACE2}
\end{center}
\end{table}

For the instances with the treewidth upper bound (best of PACE) at
most 200, our implementation is clearly superior to PACE winners.
For all of those 59 instances, our upper bound is at most one plus
the best of PACE upper bound; for 56 instances, our upper bound is
at least as good as the best of PACE; for 29 instances our upper bound is 
strictly better than the best of PACE.  
The largest improvement is 26 percent reduction in
the upper bound from 106 to 78 for instance ``he112''.

For the remaining 24 instances, the results are mixed: the best of PACE upper bound
is better than ours for 14 instance while ours are better for 8 instances.
When compared with individual winners of PACE, we may say that our 
implementation is
at least competitive even on instances in this range.

For the 17 instances with more than $100000$ vertices from PACE 2017 competition,
our implementation fails to produce meaningful results due to
either the lack of memory or the lack of time. We remark that our implementation
is not prepared for such large instances. Certainly, more work needs to be done in this respect.

\subsection*{Acknowledgments}
This work was done while the author was on leave at Utrecht University.
He thanks the Department of Information and Computing Sciences and
Hans Bodlaender for hospitality.

\end{document}